\documentstyle[12pt,a4]{article}
\input epsf

\global\arraycolsep=2pt

\begin{document}

\begin{titlepage}

\begin{flushright}
CERN-TH.7540/94\\
SHEP 95-10\\
hep-ph/9504217
\end{flushright}

\vspace{0.5cm}

\begin{center}
\Large\bf The Invisible Renormalon
\end{center}

\vspace{0.5cm}

\begin{center}
Guido Martinelli$^{a,*}$, Matthias Neubert$^a$ and Chris
T.~Sachrajda$^b$\\
\vspace{1cm}
{\sl $^a$ Theory Division, CERN, CH-1211 Geneva 23, Switzerland}\\
\vspace{0.2cm}
{\sl $^b$ Department of Physics, University of Southampton}\\
{\sl Southampton SO17 1BJ, United Kingdom}
\end{center}

\vspace{0.7cm}

\begin{abstract}
We study the structure of renormalons in the Heavy Quark Effective
Theory, by expanding the heavy quark propagator in po\-wers of
$1/m_Q$. We demonstrate that the way in which renormalons appear
depends on the regularisation scheme used to define the effective
theory. In order to investigate the relation between ultraviolet
renormalons and power divergences of matrix elements of
higher-dimensional operators in the heavy quark expansion, we perform
calculations in dimensional regularisation and in three different
cut-off regularisation schemes. In the case of the kinetic energy
operator, we find that the leading ultraviolet renormalon which
corresponds to a quadratic divergence, is absent in all but one (the
lattice) regularisation scheme. The nature of this ``invisible
renormalon'' remains unclear.
\end{abstract}

\vspace{0.7cm}

\centerline{(Submitted to Nuclear Physics B)}

\vspace{1.2cm}

\noindent
CERN-TH.7540/94\\
April 1995

\vspace{1.5cm}

\centerline{$^*$ On leave of absence from Dip.\ di Fisica,
Universit\`a degli Studi di Roma ``La Sapienza''}

\end{titlepage}

\section{Introduction}
\label{sec:intro}

Several recent papers have been devoted to the study of renormalons
in the Heavy Quark Effective Theory (HQET)
\cite{bigi}--\cite{manohar}. These studies exploit and extend the
understanding of the relevance of renormalons in field theory, and
for Operator Product Expansions (OPE) in particular, developed in
refs.~\cite{tHof}--\cite{Muel}. The present interest was initiated by
the observation that the pole mass of a heavy quark, which is an
important parameter in the heavy quark expansion, has an intrinsic
ambiguity of order $\Lambda_{\rm QCD}$ due to the presence of
infrared (IR) renormalons \cite{bigi,bb}. In addition to the infrared
renormalons and other non-perturbative effects present in QCD,
renormalon ambiguities also arise in the HQET as a result of the
expansion in powers of $1/m_Q$. A consequence of these additional
renormalons is the appearance of non-perturbative effects in the
Wilson coefficients which relate operators in the effective theory to
operators in the full theory (QCD). This problem is common to all
effective theories obtained from an expansion in inverse powers of a
large scale. Since these effects are an artefact of the construction
of the effective theory and are absent in the original theory, they
have to cancel in predictions for physical quantities. In the context
of the HQET, it has been established that the IR renormalon
ambiguities in the Wilson coefficients are cancelled by ultraviolet
(UV) renormalon ambiguities in the matrix elements of the operators
in the effective theory. How this cancellation occurs was traced
explicitly, to order $1/m_Q$ in the heavy quark expansion, for both
inclusive \cite{bbz,ns,manohar} and exclusive \cite{ns} weak decays
of hadrons containing a heavy quark. A non-perturbative approach for
the elimination of the ambiguities in the definition of the pole
mass, or more generally in the matrix elements of the operators
appearing in the HQET, has been proposed in ref.~\cite{7517}.

In the present paper we study in more detail the structure of
renormalons in the HQET. To be specific, we consider QCD Green
functions that depend on a large scale $m_Q$ (the heavy quark mass)
and a small scale $k$ (the so-called residual momentum). The purpose
of the heavy quark expansion is to disentangle the physics on
different length scales by introducing a factorisation scale
$\lambda$ such that $k\ll\lambda\ll m_Q$. Contributions from virtual
momenta above $\lambda$ are calculable in perturbation theory and
attributed to Wilson coefficients, whereas contributions from virtual
momenta below $\lambda$ are contained in the matrix elements of the
operators in the effective theory. If this program is performed with
a ``hard'' factorisation scale, these matrix elements will, for
dimensional reasons, diverge as powers of the UV cut-off $\lambda$.
The appearance of power divergences leads to non-perturbative
ambiguities, since factors such as
\begin{equation}\label{eq:arg}
   \lambda\,\exp\bigg( -{2\pi\over\beta_0\,\alpha_s(\lambda)}
   \bigg) = \Lambda_{\rm QCD}
\end{equation}
give non-vanishing contributions, which do not appear in perturbation
theory~\cite{mms}. For practical reasons, however, one usually
calculates the Wilson coefficients using dimensional regularisation.
In this case, by definition, power divergences do not appear, and the
``hard'' factorisation scale $\lambda$ is replaced by a ``soft''
renormalisation scale $\mu$. In such a scheme it is unavoidable that
the Wilson coefficients receive contributions from momenta below
$\mu$ (so-called IR renormalons), and the matrix elements receive
contributions from momenta above $\mu$ (so-called UV renormalons).
These contributions lead to a factorial growth of the coefficients in
the perturbative expansion of the Wilson coefficients and matrix
elements. The corresponding perturbative series are divergent and not
Borel summable. Since the choice of the renormalisation scale is
arbitrary, however, the effects of IR and UV renormalons must cancel
each other if one combines the perturbative series for the
coefficient functions and matrix elements.

Our main point is to demonstrate that the way in which renormalons
appear in the HQET (and, by the same argument, in other effective
field theories) is not universal, but depends on the regularisation
scheme used to define the effective theory. To this end, we
investigate in detail the relation between UV renormalons and power
divergences of hadronic matrix elements of higher-dimensional
operators. We study explicitly the $1/m_Q$ expansion of the heavy
quark propagator in different regularisation schemes: dimensional
regularisation and three different schemes with a hard UV cut-off
(Pauli--Villars, momentum flow \cite{new} and lattice
regularisation).

In general, from the degree of divergence of the matrix elements of
an operator in the HQET one can deduce the position of the UV
renormalon singularities in the Borel transform of these matrix
elements (see section 2 for a detailed explanation). By dimensional
arguments we expect that the kinetic energy operator for a heavy
quark is quadratically divergent, since it can mix with the identity
operator under renormalisation. Indeed, this mixing has been observed
in lattice perturbation theory\cite{mms}. The corresponding
renormalon singularity is absent, however.\footnote{The absence of
the corresponding IR renormalon in the pole mass was previously noted
in ref.~\protect\cite{bb}.} This ``missing'' singularity is what we
have called the ``invisible renormalon''. Even more surprising is the
absence, at one-loop order, of the quadratic divergence in the two
other schemes considered in our study, i.e.\ the Pauli-Villars and
momentum flow regularisations. We have been unable to understand
whether there is a symmetry, broken by the lattice regularisation,
which prevents the appearance of the quadratic divergence in the
other cases, or whether this is an artefact of one-loop perturbation
theory. In the latter case, the quadratic divergence would appear at
higher orders. Given the relevance of the kinetic energy operator for
heavy hadron spectroscopy and inclusive decays
\cite{spectrum}--\cite{inclu4}, an understanding of this puzzle is
very important.

The remainder of this paper is organised as follows: In
sect.~\ref{sec:2} we summarise and discuss the main results of our
study. Technical details and explicit calculations in different
regularisation schemes are presented in
sects.~\ref{sec:exp}--\ref{sec:cutoff}. In sect.~\ref{sec:6} we give
our conclusions.

\section{Renormalons and power divergences}
\label{sec:2}

Perturbative expansions in QCD are asymptotic and, in general, not
Borel sum\-mable. When one tries to resum a perturbative series to
all orders, one encounters ambiguities, which indicate that
perturbation theory is by itself incomplete and must be supplemented
by non-perturbative corrections. A convenient way to analyse these
ambiguities is to consider the Borel transform $\widetilde S(u)$ of a
series $S(\alpha_s(\mu))$ with respect to the coupling constant
\cite{tHof}. Formally, the Borel sum of the series can be defined by
the integral\footnote{We follow the notations and definitions of
ref.~\protect\cite{ns}.}
\begin{equation}\label{BTinv}
   S(\alpha_s(\mu)) = \int\limits_0^\infty\!{\rm d}u\,\exp\bigg(
   -{4\pi u\over\beta_0\alpha_s(\mu)} \bigg)\,\widetilde S(u) \,,
\end{equation}
where $\beta_0=11-\frac{2}{3}\,n_f$ is the first coefficient of the
$\beta$-function. However, if the Borel transform contains
singularities on the integration contour, the result of the
integration depends on the regularisation prescription, and
$S(\alpha_s(\mu))$ is not uniquely defined in terms of $\widetilde
S(u)$. In QCD, one source of such singularities are the
higher-order diagrams in which a virtual gluon line with momentum $k$
is dressed by a number of fermion, gluon and ghost loops. More
precisely, one has to consider a gauge-invariant generalisation of
such diagrams, which is called a renormalon chain. Effectively, this
introduces the running coupling constant $\alpha_s(k)$ at the
vertices. When $\alpha_s(k)$ is expressed in terms of the coupling
constant renormalised at a fixed scale $\mu$, the appearance of
powers of large logarithms leads to a factorial divergence in the
expansion coefficients of the perturbative series. Associated with
this are renormalon singularities in the Borel transform $\widetilde
S(u)$. Depending on whether these singularities are related to the
region of large or small virtual momenta, they are referred to as UV
or IR renormalons \cite{tHof}--\cite{Muel}.

Although the resummation of renormalon chains corresponds to only a
partial resummation of the perturbative series, it elucidates many
non-perturbative effects and thus provides an interesting,
non-trivial approximation. If one accepts this so-called ``bubble
approximation'', the renormalon singularities occur as poles on the
real axis in the Borel plane.\footnote{When the calculations are
extended beyond the ``bubble approximation'', the poles become
replaced by branch points of cut singularities.} Let us denote the
position of the nearest pole on the positive $u$-axis by $u_0$, i.e.\
$\widetilde S(u)=r_0/(u-u_0) + \mbox{terms that are regular at
$u=u_0$}$. Then a measure of the ambiguity in the Borel integral
(\ref{BTinv}) is given by
\begin{equation}\label{DeltaS}
   \Delta S = r_0\,\exp\bigg(
   -{4\pi u_0\over\beta_0\,\alpha_s(\mu)} \bigg)
   = r_0\,\bigg( {\Lambda_{\rm QCD}\over\mu} \bigg)^{2 u_0} \,.
\end{equation}
In the last step we have used the one-loop expression for the
running coupling constant.

As mentioned above, in dimensional regularisation power divergences
are hidden because of the absence of an intrinsic mass scale in the
computation. However, perturbation theory ``knows'' about these
divergences in the form of renormalon singularities in the Borel
plane. In fact, there is a one-to-one correspondence between the
structure of the UV renormalon poles in dimensional regularisation
and the power divergences in regularisation schemes with a hard
cut-off. For instance, if a quantity is linearly divergent in
one-loop perturbation theory, then its Borel transform in the bubble
approximation is logarithmically divergent at $u=1/2$, corresponding
to a pole singularity at this point. A similar argument is expected
to hold for quadratic or higher-order power divergences. To
illustrate this point we present results for the renormalised inverse
heavy quark propagator $S_{\rm eff}^{-1}(k)$ in the HQET. Here
$k=p_Q-m_Q v$ is the residual momentum, which stays finite in the
limit $m_Q\to\infty$, and $v$ is a four-velocity vector ($v^2=1$),
which is usually identified with the velocity of the hadron
containing the heavy quark. In dimensional regularisation, renormalon
effects appear as singularities in the Borel transform $\widetilde
S_{\rm eff}^{-1}(u)$. In the bubble approximation one
finds\footnote{In this paper we define the running coupling constant
in the so-called V scheme \protect\cite{BLM}, in which the one-loop
counterterms differ from those of the MS scheme by an additive
constant. Otherwise, the Borel transform has to be multiplied by a
scheme-dependent factor $e^{-C u}$.} \cite{bb}
\begin{equation}\label{Seff}
   \widetilde S_{\rm eff}^{-1}(k,u) = v\cdot k\,\bigg\{ \delta(u)
   + {6 C_F\over\beta_0}\,\bigg[ \bigg( {-2 v\cdot k\over\mu}
   \bigg)^{-2u}\,{\Gamma(1-u)\,\Gamma(-1+2u)\over\Gamma(2+u)}
   + {1\over 2 u} - R_Z(u) \bigg] \bigg\}
\end{equation}
in the Landau gauge. The scheme-dependent function $R_Z(u)$, which
for renormalisation schemes with analytic counterterms (such as
MS-like schemes) is entire in the complex $u$-plane, is irrelevant
for our purposes. Expanding the above expression around $u=0$ and
inserting the result into the Borel integral (\ref{BTinv}), one finds
\begin{equation}\label{Sdr}
   S_{\rm eff}^{-1}(k) = v\cdot k\,\bigg\{ 1
   + {C_F\alpha_s\over\pi}\,\bigg( {3\over2}\,
   \ln{-2 v\cdot k\over\mu} + \mbox{const.} \bigg)
   + O(\alpha_s^2) \bigg\} \,,
\end{equation}
where $C_F=4/3$ is a colour factor. The $\Gamma$-functions in the
numerator of (\ref{Seff}) define the positions of the renormalon
singularities. The UV renormalons are given by the poles of
$\Gamma(-1+2 u)$ and are thus located at
$u=\frac{1}{2},0,-\frac{1}{2},\dots$\footnote{The pole at $u=0$ is
removed by renormalisation, however.} They are due to the
contributions of virtual momenta above $\mu$ in loop integrals. Note
that there appear also IR renormalons (at $u=1,2,\dots$), which are
due to the contributions of virtual momenta below the soft scale $k$.
Since, by construction, the effective theory and the full theory
describe the same dynamics in the IR region, these renormalons
reflect truly non-perturbative effects of QCD and are thus
independent of the UV regularisation scheme. They are not related to
the construction of the HQET and are therefore not the topic of our
discussion. The UV renormalon at $u=1/2$ is the nearest singularity
on the integration contour of the Borel integral in (\ref{BTinv}). An
expansion of the Borel transform around this point leads to
\begin{equation}\label{Suresid}
   \widetilde S_{\rm eff}^{-1}(k,u) = -{2 C_F\over\beta_0}\,
   {\mu\over u-{1\over 2}} + \dots \,.
\end{equation}
The linear dependence on $\mu$ reflects the linear UV divergence of
the inverse propagator in the effective theory, which in dimensional
regularisation is not seen in perturbation theory. Using the above
result together with (\ref{DeltaS}), we find that the renormalon
ambiguity in the definition of the inverse propagator is of order
$\Lambda_{\rm QCD}$.

Let us now compare these results to the one-loop calculation of the
inverse propagator performed with a hard Pauli--Villars cut-off
$\lambda$. In this case one finds (for details see
sect.~\ref{sec:cutoff})
\begin{equation}\label{SPV}
   S_{\rm eff}^{-1}(k) = -{C_F\alpha_s\over 2}\,\lambda
   + v\cdot k\,\bigg\{ 1 + {C_F\alpha_s\over\pi}\,\bigg(
   {3-a\over 2}\,\ln{-2 v\cdot k\over\lambda}
   + \mbox{const.} \bigg) \bigg\} + O(\alpha_s^2) \,.
\end{equation}
We have given the result for an arbitrary covariant gauge. The Landau
gauge corresponds to $a=0$, whereas the Feynman gauge corresponds to
$a=1$. Note that the coefficient of the logarithmic term is the same
in (\ref{Sdr}) and (\ref{SPV}). The linearly UV divergent term in
(\ref{SPV}) is in correspondence with the term proportional to $\mu$
in (\ref{Suresid}), which leads to a renormalon ambiguity of order
$\Lambda_{\rm QCD}$. Note that in the case of the cut-off
regularisation the power divergence leads to an ambiguity of the same
order, as explained above [see (\ref{eq:arg})].

This correspondence between UV renormalons and power divergences
works in several cases. However, an apparent exception is provided by
the quark matrix element of the kinetic energy operator $\bar h_v\,(i
D_\perp)^2 h_v$, which appears at order $1/m_Q$ in the expansion of
the inverse propagator.\footnote{Here $h_v$ denotes the
velocity-dependent heavy quark field in the HQET, and
$D_{\perp,\mu}=D_\mu -v\!\cdot\!D\,v_\mu$ contains the ``spatial''
components of the covariant derivative. For details see
ref.~\protect\cite{review} and references therein.} This
dimension-five operator has the same quantum numbers as the
lower-dimen\-sional operators $\bar h_v\,h_v$ (dimension-three) and
$\bar h_v\,i v\!\cdot\!D\,h_v$ (dimension-four). Hence, in a generic
regularisation scheme with a dimensionful regulator, one expects a
mixing with these operators, leading to quadratic and linear UV
divergences. Indeed, these divergences appear in the lattice
formulation of the HQET, where, at one-loop order, one
finds\footnote{ The calculation was originally performed for a heavy
quark at rest. The generalisation to an arbitrary four-velocity is
possible, although there are considerable subtleties in formulating
the effective theory at non-zero velocity in Euclidean space
\cite{mandula,ugo}.} \cite{mms}
\begin{equation}\label{kinlat}
   S_{\rm kin}^{-1}(k) = \langle\,k\,|\,\bar h_{v}\,(i D_\perp)^2
   h_{v}\,|\,k\,\rangle = {W\,\alpha_s\over a^2}
   + {2 X\,\alpha_s\over a}\,v\cdot k + \dots \,.
\end{equation}
Here $W$ and $X$ are dimensionless constants, and the UV cut-off is
provided by the inverse lattice spacing, $\lambda=1/a$. The puzzle
that we wish to point out is that the quadratic divergence is {\it
not\/} seen in some other cut-off regularisation schemes (such as
Pauli--Villars). Moreover, in the bubble approximation we do not find
an UV renormalon pole corresponding to the quadratic divergence. The
nearest UV singularity in the Borel transform of $S_{\rm
kin}^{-1}(k)$ is located at $u=\frac{1}{2}$, and an expansion around
this point gives
\begin{equation}\label{kinBor}
    \widetilde S_{\rm kin}^{-1}(k,u) = {6 C_F\over\beta_0}\,
    {\mu\over u-{1\over 2}}\,v\cdot k + \dots \,.
\end{equation}
This behaviour corresponds to the linearly divergent term in
(\ref{kinlat}). However, we do not find a term of the form
$\mu^2/(u-1)$, which would correspond to the quadratic divergence. We
shall refer to this ``missing'' renormalon pole at $u=1$ as the {\it
invisible renormalon}.

There are several possible explanations for this puzzle. The general
argument given above is that there is a relation between the degree
of divergence of the matrix elements of higher-dimensional operators
and the position of the singularities in their Borel transform in
dimensional regularisation. The coefficients of the divergent terms
and of the renormalon singularities, however, are not universal and
may be zero in some regularisation schemes. Indeed, there is no
reason why renormalon singularities or power divergences should be
the same for all versions of the theory, since the ambiguities
related to these renormalons (or powers divergences) are spurious and
cancel in the predictions for physical quantities. It is therefore
conceivable to find some regularisation schemes in which at one-loop
order the coefficient of the quadratic divergence of the kinetic
operator happens to be zero.\footnote{If this is the case, a
quadratic divergence is likely to appear at two-loop order, since at
one-loop order the operator $\bar h_v\,(i D_\perp)^2 h_v$ mixes with
$\bar h_v\,iv\!\cdot\!D\,h_v$, which itself mixes with $\bar
h_v\,h_v$.} However, it is surprising to us that we could find only a
single regularisation scheme in which this coefficient {\it does not
vanish}. Moreover, even if one accepts this point of view one still
has to explain the absence of a renormalon pole at $u=1$ in the Borel
transform (\ref{kinBor}). A possibility would be that such a
singularity appears only when one goes beyond the bubble
approximation.

On the other hand, one may wonder whether the appearance of a
quadratic divergence in the lattice regularisation scheme is due to
the breaking of some symmetry preserved in other regularisation
schemes. In this case the invisible renormalon would be nothing more
than a lattice artefact. Examples of such broken symmetries are
Lorentz symmetry and, as a consequence, the so-called
reparametrisation invariance of the HQET \cite{luma,cino}. We note
that the breaking of reparametrisation invariance is responsible for
the fact that the renormalisation constants of the operators $\bar
h_v\,i v\cdot D\,h_v$ and $\bar h_v\,(i D_\perp)^2 h_v$ differ by a
finite term of order $\alpha_s$ in the lattice version of the HQET
\cite{mms}, whereas they are the same in the continuum version of the
theory \cite{luma}. However, reparametrisation invariance does not
forbid the mixing between the operators $\bar h_v\,(i D_\perp)^2 h_v$
and $\bar h_v\,h_v$, which is responsible for the quadratic
divergence.

This concludes our main discussion. In the following sections we
present the details of the calculations, whose results were referred
to above.

\section{Expansion of the inverse propagator}
\label{sec:exp}

Let us investigate the appearance of renormalons in the $1/m_Q$
expansion of the inverse heavy quark propagator, which is the
simplest Green function for a heavy quark. We shall generalise the
analysis of Beneke and Braun \cite{bb} by including terms of order
$1/m_Q$. We work in dimensional regularisation and adopt the bubble
approximation to investigate the singularities in the Borel plane.
For technical details of the calculation the reader is referred to
refs.~\cite{bb,ns,Bene}.

We start from the quark propagator in the full theory,
\begin{equation}\label{ftp}
   S(p,m) = {1\over\,\rlap/\!p - m - \Sigma(p,m)} \,,
\end{equation}
where $m$ denotes the bare mass, and
\begin{equation}
   \Sigma(p,m) = m\,\Sigma_1(p^2,m^2)
   + (\,\rlap/\!p - m)\,\Sigma_2(p^2,m^2)
\end{equation}
is the self-energy. We write the heavy quark momentum as $p=m_Q v+k$,
where $m_Q$ is the expansion parameter of the HQET (in general
$m_Q\ne m$), and consider $\Sigma_{1,2}\equiv\Sigma_{1,2}(k,m)$ as
functions of the residual momentum $k$. Next we define a projected
propagator $S_P(k,m_Q)$ from the relation\footnote{The authors of
ref.~\protect\cite{bb} have projected the inverse propagator instead
of the propagator itself. This procedure is not applicable if one
wants to include $1/m_Q$ corrections in the HQET.}
\begin{equation}
  {1+\rlap/v\over 2}\,S_P(k,m_Q)\equiv
  {1+\rlap/v\over 2}\,S(p,m)\,{1+\rlap/v\over 2} \,.
\end{equation}
Our goal is to construct the heavy quark expansion of
$S_P^{-1}(k,m_Q)$. Including terms of order $1/m_Q$, we find
\begin{equation}
   S_P^{-1}(k,m_Q) = (1-\Sigma_2)\,\bigg\{ (m_Q - m) + v\cdot k
   + {k_\perp^2\over 2 m_Q}\,\Big( 1 - \delta/2 \Big)^{-1}
   \bigg\} - \Sigma_1\,m + O(1/m_Q^2) \,,
\end{equation}
where $k_\perp^2=k^2-(v\cdot k)^2$, and
\begin{equation}
   \delta = 1 - {(1+\Sigma_1-\Sigma_2)\,m\over(1-\Sigma_2)\,m_Q} \,.
\end{equation}
The result simplifies if we restrict ourselves to the bubble
approximation, which corresponds to the first term in an expansion in
powers of $1/\beta_0$. We can then use the fact that $\Sigma_1$,
$\Sigma_2$, $\delta$ and $(m_Q-m)$ are of order $1/\beta_0$. At the
same time we can substitute the bare mass $m$ by the HQET expansion
parameter $m_Q$ in the expressions for $\Sigma_i$. This leads to
\begin{eqnarray}
   S_P^{-1}(k,m_Q) &=& -\delta m - \Big[ \Sigma_1(k,m_Q)
    - \Sigma_1(0,m_Q) \Big]\,m_Q \nonumber\\
   &&\mbox{}+ \Big[ 1 - \Sigma_2(k,m_Q) \Big]\,
    \bigg( v\cdot k + {k_\perp^2\over 2 m_Q} \bigg)
    + O(1/\beta_0^2,1/m_Q^2) \,,
\end{eqnarray}
where $\delta m=m_{\rm pole}-m_Q$, and
\begin{equation}
   m_{\rm pole} = m\,\Big\{ 1 + \Sigma_1(0,m_Q) \Big\}
   + O(1/\beta_0^2)
\end{equation}
is the pole mass. The heavy quark expansion is consistent as long as
the so-called residual mass $\delta m$ is a parameter of order
$\Lambda_{\rm QCD}$ \cite{FNL}.

In order to see the appearance of renormalons, one has to consider
the Borel transform of the inverse propagator. In the bubble
approximation, explicit expressions for Borel transforms of the pole
mass and the functions $\Sigma_i(k,m_Q)$ have been derived by Beneke
and Braun \cite{bb}. By expanding their results in powers of $1/m_Q$,
we obtain
\begin{eqnarray}\label{eq:bt}
   \widetilde S_P^{-1}(k,m_Q,u) &=& - \delta\widetilde m(u)
    + \delta(u)\,\bigg( v\cdot k + {k_\perp^2\over 2 m_Q} \bigg)
    \nonumber\\
   &&\mbox{}+ {6 C_F\over\beta_0}\,\Bigg\{
    v\cdot k\,\Big[ (1-u^2)\,R_1 + R_2 \Big] \nonumber\\
   &&\phantom{ +{6 C_F\over\beta_0}\,\Bigg\{ }
    \mbox{}+ {k_\perp^2\over 2 m_Q}\,\Big[ (1-u^2)\,R_1
    + (1-2 u)\,R_2 \Big] \\
   &&\phantom{ +{6 C_F\over\beta_0}\,\Bigg\{ }
    \mbox{}+ {(v\cdot k)^2\over 2 m_Q}\,\Big[
    X(u)\,R_1 - 3\,R_2 \Big] \Bigg\} + O(1/\beta_0^2,1/m_Q^2) \,,
    \nonumber
\end{eqnarray}
where
\begin{eqnarray}\label{abbrev}
   \delta\widetilde m(u) &=& \widetilde m_{\rm pole}(u)
    - m_Q\,\delta(u) \,,\nonumber\\
   \widetilde m_{\rm pole}(u) &=& m\,\bigg\{ \delta(u)
    + {6 C_F\over\beta_0}\,\bigg[ (1-u)\,R_1 - {1\over 2 u}
    + R_m(u) \bigg] \bigg\} \,,\nonumber\\
   R_1 &=& \bigg( {m_Q\over\mu} \bigg)^{-2u}\,
    {\Gamma(u)\,\Gamma(1-2u)\over\Gamma(3-u)} \,,\nonumber\\
   R_2 &=& \bigg( {-2 v\cdot k\over\mu} \bigg)^{-2u}\,
    {\Gamma(1-u)\,\Gamma(-1+2u)\over\Gamma(2+u)} \,,\nonumber\\
   X(u) &=& - {6 + 5u - 2u^2 - 8u^3 - 4u^4\over 2(1+2u)} \,.
\end{eqnarray}
The scheme-dependent function $R_m(u)$ is irrelevant to our
discussion. It is convenient to rewrite (\ref{eq:bt}) in the form of
a convolution of Borel transforms, which is defined as
\begin{equation}
   \widetilde f(u) * \widetilde g(u) \equiv \widetilde{f\cdot g}(u)
   = \int\limits_0^u\!{\rm d}u'\,\widetilde f(u')\,
   \widetilde g(u-u') \,.
\end{equation}
Then the result takes the form
\begin{eqnarray}\label{nice}
   \widetilde S_P^{-1}(k,m_Q,u) &=& -\delta\widetilde m(u)
    + \widetilde Z_Q^{-1}(m_Q,u) * \bigg\{
    \widetilde S_{\rm eff}^{-1}(k,u) \nonumber\\
   &&\mbox{}+ {1\over 2 m_Q}\,\Big[
    \widetilde C_{\rm kin}(m_Q,u) * \widetilde S_{\rm kin}^{-1}(k,u)
    + \widetilde C_{(v\cdot D)^2}(m_Q,u)
    * \widetilde S_{(v\cdot D)^2}^{-1}(k,u) \Big] \nonumber\\
   &&\mbox{}+ O(1/m_Q^2) \bigg\} \,,
\end{eqnarray}
in which the heavy quark expansion is explicitly realised, i.e.\ the
dependence on  the two scales $m_Q$ and $k$ is disentangled. The
dependence on the heavy quark mass $m_Q$ is contained in the
functions
\begin{eqnarray}\label{Zfactors}
   \widetilde Z_Q^{-1}(m_Q,u) &=& \delta(u)
    + {6 C_F\over\beta_0}\,\bigg[ (1-u^2)\,R_1 - {1\over 2 u}
    + R_Z(u) \bigg] \,,\nonumber\\
   \phantom{ \bigg[ }
   \widetilde C_{\rm kin}(m_Q,u) &=& \delta(u) \,,\nonumber\\
   \widetilde C_{(v\cdot D)^2}(m_Q,u) &=& {6 C_F\over\beta_0}\,
    \bigg[ X(u)\,R_1 + {3\over 2 u} + R_C(u) \bigg] \,,
\end{eqnarray}
which are the Borel transforms (in the bubble approximation) of the
wave-function renormalisation factor $Z_Q^{-1}$ and the Wilson
coefficients of the operators $\bar h_v\,(i D_\perp)^2 h_v$ (kinetic
energy operator) and $\bar h_v\,(i v\!\cdot\!D)^2 h_v$ in the
effective
Lagrangian of the HQET \cite{EiH1,FGL}. Note that $\widetilde C_{\rm
kin}=\delta(u)$ implies $C_{\rm kin}=1$, as required by
reparametrisation invariance \cite{luma}. The dependence on the small
scale $k$ in (\ref{nice}) resides in the quantities
\begin{eqnarray}\label{Sfuns}
   \widetilde S_{\rm eff}^{-1}(k,u) &=& v\cdot k\,\bigg\{
    \delta(u) + {6 C_F\over\beta_0}\,\bigg[ R_2 + {1\over 2 u}
    - R_Z(u) \bigg] \bigg\} \,,\nonumber\\
   \widetilde S_{\rm kin}^{-1}(k,u) &=& k_\perp^2\,\bigg\{
    \delta(u) + {6 C_F\over\beta_0}\,\bigg[ (1-2u)\,R_2
    + {1\over 2 u} - R_Z(u) \bigg] \bigg\} \nonumber\\
   &&\mbox{}+ (v\cdot k)^2\,{6 C_F\over\beta_0}\,\bigg[
    -3 R_2 - {3\over 2 u} - R_C(u) \bigg] \,,\nonumber\\
   \widetilde S_{(v\cdot D)^2}^{-1}(k,u) &=& (v\cdot k)^2\,\delta(u)
    + O(1/\beta_0) \,,
\end{eqnarray}
which are the Borel transforms (in the bubble approximation) of the
quark matrix elements of the operators appearing in the effective
Lagrangian of the HQET. Their calculation will be outlined in
sect.~\ref{sec:3}. Note that it is sufficient to keep terms of order
unity in $\widetilde S_{(v\cdot D)^2}$, since the corresponding
Wilson coefficient $\widetilde C_{(v\cdot D)^2}$ in (\ref{Zfactors})
starts at order $1/\beta_0$. The functions $R_Z(u)$ and $R_C(u)$ in
(\ref{Zfactors}) and (\ref{Sfuns}) depend on the renormalisation
scheme. They are irrelevant to our discussion.

It is instructive to trace how renormalons are introduced in the
construction of the heavy quark expansion, and in which way they
cancel between the coefficient functions and matrix elements. To
start with, we note that even in the full theory the self-energy
contains IR renormalons from the contributions of virtual momenta
below the scale $k$. Their positions are determined by the factor
$\Gamma(1-u)$ contained in $R_2$ in (\ref{abbrev}). Note that the
leading IR renormalon at $u=1$ is not forbidden by gauge invariance,
since the self-energy is not gauge invariant and there exists a
gauge-variant operator of dimension two \cite{bb}. What is introduced
in the process of constructing the heavy quark expansion are new
renormalon singularities at half-integer values of $u$. In the
coefficient functions there appear IR renormalons at positions
$u=\frac{1}{2},1,\frac{3}{2},\dots$ determined by the factor
$\Gamma(1-2 u)$ contained in $R_1$. Likewise, in the matrix elements
there appear UV renormalons at positions
$u=\frac{1}{2},0,-\frac{1}{2},\dots$ determined by the factor
$\Gamma(-1+2 u)$ contained in $R_2$. Since these singularities are
absent in the original theory, they must cancel, order by order in
$1/m_Q$, in the effective theory. Consider now our results
(\ref{eq:bt}) and (\ref{nice}) to see how the IR and UV renormalon
poles conspire. At order $1/m_Q$ only the renormalons at $u=1/2$ need
to be considered. We observe that the IR renormalon in
$\delta\widetilde m$ matches with the UV renormalon in $\widetilde
S_{\rm eff}^{-1}$, and the IR renormalon in the coefficient
$\widetilde Z_Q^{-1}$ which multiplies $\widetilde S_{\rm eff}^{-1}$
matches with the UV renormalon in $\widetilde S_{\rm kin}^{-1}$. This
becomes explicit if we expand (\ref{eq:bt}) around $u=1/2$:
\begin{eqnarray}
   {\beta_0\over 8 C_F}\,\widetilde S_P^{-1}(k,m_Q,u)
   &=& - m_Q\,{1\over 2(1-2 u)}\,{\mu\over m_Q} \nonumber\\
    &&\mbox{}+ v\cdot k\,\bigg[ {3\over 4(1-2 u)}\,{\mu\over m_Q}
    - {1\over 2(2 u-1)}\,{\mu\over v\cdot k} \bigg] \nonumber\\
   &&\mbox{}+ {(v\cdot k)^2\over 2 m_Q}\,{3\over 2(2 u-1)}\,
    {\mu\over v\cdot k} + \dots \,,
\end{eqnarray}
where the ellipses represent terms that are regular at
$u=\frac{1}{2}$. Notice the absence of an UV renormalon at
$u=\frac{1}{2}$ in the term proportional to $k_\perp^2$ in
$\widetilde S_{\rm kin}^{-1}$ given in (\ref{Sfuns}). Such a
renormalon is forbidden, since it would give rise to a non-local
behaviour of the form $\mu\,k_\perp^2/v\cdot k$. On the other hand,
nothing forbids an IR renormalon at $u=1$ in $\delta\widetilde m$,
which could conspire with an UV renormalon at $u=1$ in the term
proportional to $(v\cdot k)^2$ in $\widetilde S_{\rm
kin}^{-1}$.\footnote{There is in fact a pole at $u=1$ in $\widetilde
S_{\rm kin}^{-1}$, but it has nothing to do with the UV region.
Computing the matrix element of $\bar h_v\,(i D_\perp)^2 h_v$ with a
hard IR cut-off we find that this pole disappears. Hence, it is
related to the region of momenta below $k$. This IR renormalon pole
appears also in the full theory.} The fact that this does not appear
is the puzzle mentioned in the introduction.

Before we proceed, let us note that from an expansion of the
expressions in (\ref{Zfactors}) and (\ref{Sfuns}) around $u=0$ one
can extract the one-loop expressions for the corresponding
quantities. Keeping only logarithmic terms, we find
\begin{eqnarray}
   Z_Q^{-1}(m_Q) &=& 1 - {3-a\over 2}\,{C_F\alpha_s\over\pi}\,
    \ln{m_Q\over\mu} \,,\nonumber\\
   C_{(v\cdot D)^2}(m_Q) &=& {3(3-a)\over 2}\,
    {C_F\alpha_s\over\pi}\,\ln{m_Q\over\mu} \,,\nonumber\\
   S_{\rm eff}^{-1}(k) &=& v\cdot k\,\bigg\{ 1 + {3-a\over 2}\,
    {C_F\alpha_s\over\pi}\,\ln{-2 v\cdot k\over\mu} \bigg\}
    \,, \nonumber\\
   S_{\rm kin}^{-1}(k) &=& k_\perp^2\,\bigg\{ 1 + {3-a\over 2}\,
    {C_F\alpha_s\over\pi}\,\ln{-2 v\cdot k\over\mu} \bigg\}
    \nonumber\\
   &&\mbox{}- {3(3-a)\over 2}\,{C_F\alpha_s\over\pi}\,(v\cdot k)^2\,
    \ln{-2 v\cdot k\over\mu} \,,
\end{eqnarray}
where we have given the results for an arbitrary covariant gauge. The
expression for the Wilson coefficient $C_{(v\cdot D)^2}$ has been
derived (in the Feynman gauge) in ref.~\cite{FGL}.

\section{Renormalons in HQET matrix elements}
\label{sec:3}

In this section we show in more detail that the functions in
(\ref{Sfuns}) can be identified with the Borel transforms of quark
matrix elements of operators in the effective Lagrangian of the HQET.
At order $1/m_Q$, this Lagrangian reads \cite{EiH1,FGL}
\begin{eqnarray}
   {\cal{L}}_{\rm HQET} &=& \bar h_v\,i v\!\cdot\!D\,h_v
    + {1\over 2 m_Q}\,\Big[ C_{\rm kin}\,\bar h_v\,
    (i D_\perp)^2 h_v + C_{(v\cdot D)^2}\,
    \bar h_v\,(i v\!\cdot\!D)^2 h_v \nonumber\\
   &&\phantom{ \bar h_v\,i v\!\cdot\!D\,h_v + {1\over 2 m_Q}\,\Big[ }
    + C_{\rm mag}\,{g_s\over 2}\,
    \bar h_v\,\sigma_{\mu\nu} G^{\mu\nu} h_v \Big] + O(1/m_Q^2) \,,
\end{eqnarray}
with $C_{\rm kin}=1$ by reparametrisation invariance \cite{luma}. The
so-called chromo-magnetic operator $\bar h_v\,\sigma_{\mu\nu}
G^{\mu\nu} h_v$ plays no role for our discussion here, since its
matrix element between quark states vanishes. We will now outline the
calculation of the matrix elements of the other operators between
heavy quark states with velocity $v$ and residual momentum $k$, using
the Landau gauge.

The Borel transform of the inverse heavy quark propagator in the
effective theory has been calculated by Beneke and Braun \cite{bb}.
The result is
\begin{equation}\label{Seffbb}
   S_{\rm eff}^{-1}(k) = \langle\,k\,|\,\bar h_v\,i
v\!\cdot\!D\,h_v\,
   |\,k\,\rangle ~\stackrel{\rm B.T.}{\to}~
   v\cdot k\,\bigg\{ \delta(u) + {6 C_F\over\beta_0}\,R_2 \bigg\} \,.
\end{equation}
After UV renormalisation, which amounts to removing the pole at $u=0$
contained in $R_2$, this leads to the expression for $\widetilde
S_{\rm eff}(k,u)$ given in (\ref{Sfuns}).

\begin{figure}[h]
   \epsfxsize=12cm
   \centerline{\epsffile{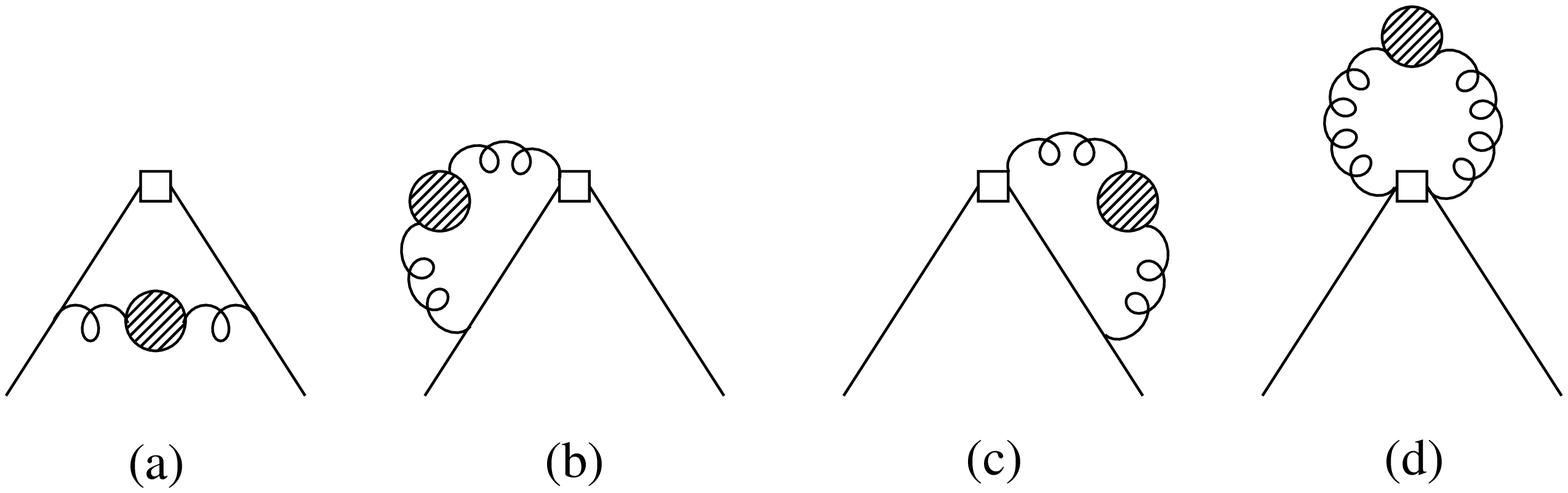}}
   \centerline{\parbox{13cm}{\caption{\label{fig:1}
Diagrams contributing to the quark matrix element of the kinetic
energy operator (indicated by a square).}}}
\end{figure}

Let us now turn to the calculation of the Borel transform of the
matrix element of the kinetic energy operator $\bar h_v\,(i
D_\perp)^2 h_v$. The relevant diagrams are depicted in
fig.~\ref{fig:1}. The shaded bubble represents the Borel transform
of the resummed gluon propagator in the Landau gauge. In the bubble
approximation, the Borel transform of any one-loop diagram is simply
obtained by using this propagator instead of the usual one
\cite{bb,Bene}. We find that the contribution of the ``vertex
diagram'' shown in fig.~\ref{fig:1}(a) is
\begin{equation}
   {6 C_F\over\beta_0}\,R_2\,\Big[ (1-2u)\,k_\perp^2
   - 5\,(v\cdot k)^2 \Big] \,.
\end{equation}
The two ``sail diagrams'' depicted in fig.~\ref{fig:1}(b) and (c),
which have a gluon attached to the operator insertion, each give a
contribution
\begin{equation}
   {6 C_F\over\beta_0}\,R_2\,(v\cdot k)^2 \,.
\end{equation}
In dimensional regularisation the tadpole diagram of
fig.~\ref{fig:1}(d) vanishes. Adding the tree-level contribution to
the above results, we obtain
\begin{equation}
   S_{\rm kin}^{-1}(k) = \langle\,k\,|\,\bar h_v\,(i D_\perp)^2
   h_v\,|\,k\,\rangle ~\stackrel{\rm B.T.}{\to}~
   k_\perp^2\,\delta(u) + {6 C_F\over\beta_0}\,R_2\,\Big[
   (1-2u)\,k_\perp^2 - 3\,(v\cdot k)^2 \Big] \,,
\end{equation}
in agreement with the expression for $\widetilde S_{\rm
kin}^{-1}(k,u)$ given in (\ref{Sfuns}).

Repeating the same calculation for the operator $\bar h_v\,(i v\cdot
D)^2 h_v$, we find that the vertex diagram vanishes, and only the
sail diagrams give a non-vanishing contribution. The result is
\begin{equation}
   S_{(v\cdot D)^2}^{-1}(k) = \langle\,k\,|\,\bar h_v\,
   (i v\!\cdot\!D)^2 h_v\,|\,k\,\rangle ~\stackrel{\rm B.T.}{\to}~
   (v\cdot k)^2\,\bigg\{ \delta(u) + {12 C_F\over\beta_0}\,R_2
   \bigg\} \,,
\end{equation}
which generalises the expression given in (\ref{Sfuns}) to order
$1/\beta_0$.

\section{Power divergences in HQET matrix elements}
\label{sec:cutoff}

In this section we present the results of several one-loop
calculations of the HQET matrix elements $S_{\rm eff}^{-1}(k)$ and
$S_{\rm kin}^{-1}(k)$ in regularisation schemes with a hard UV
cut-off $\lambda$. The aim is to study the correspondence between the
UV renormalon poles encountered in the previous section and the power
divergences associated with the use of a dimensionful regulator. For
simplicity, we shall perform the calculations in gauges suitable for
the regularisation method of choice. We note that the structure of
the power divergences is gauge invariant.

\subsection{Pauli--Villars regularisation}

In the simplest version of the Pauli--Villars regularisation scheme,
one substitutes for the gluon propagator in the Feynman gauge the
expression
\begin{equation}\label{eq:gpv1}
   G_{\mu\nu}(q) = -i g_{\mu\nu}\,\bigg( {1\over q^2}
   - {1\over q^2 - \lambda^2} \bigg)
   = {i g_{\mu\nu}\lambda^2\over q^2(q^2 - \lambda^2)} \,.
\end{equation}
Computing the one-loop self-energy in the effective theory and adding
the tree-level expression for the inverse propagator, we find
(assuming $v\cdot k<0$)
\begin{eqnarray}\label{SeffPV}
   S_{\rm eff}^{-1}(k) &=& v\cdot k\,\bigg\{ 1
    + {C_F\alpha_s\over\pi}\,\bigg( \ln{-2 v\cdot k\over\lambda}
    + \sqrt{x^2-1}\,\arctan\sqrt{x^2-1} \bigg) \bigg\} \nonumber\\
   &=& -{C_F\alpha_s\over 2}\,\lambda + v\cdot k\,\bigg\{ 1
    + {C_F\alpha_s\over\pi}\,\bigg( \ln{-2 v\cdot k\over\lambda}
    - 1 \bigg) \bigg\} + O(1/\lambda) \,,
\end{eqnarray}
where $x=\lambda/(-v\cdot k)$. The generalisation of this result to
an arbitrary covariant gauge has been given in (\ref{SPV}). The
linear divergence corresponds to the UV renormalon pole at
$u=\frac{1}{2}$ in (\ref{Seffbb}). It is instructive to rewrite the
result in a form that makes explicit the mixing of operators under
renormalisation. We define
\begin{eqnarray}\label{eq:pv2}
   S_{\rm eff}^{-1}(k) &=& \langle\,k\,|\,\bar h_v\,i v\!\cdot\!D\,
    h_v\,|\,k\,\rangle \nonumber\\
   \phantom{ \bigg[ }
   &=& Z_{\rm eff}^{-1}\,
    \langle\,k\,|\,\bar h_v\,i v\!\cdot\!D\,h_v\,|\,k\,\rangle_0
    + Z_{v \cdot D\to \hat  1}^{-1}\,\lambda\,\langle\,k\,|\,\bar
    h_v\,h_v\, |\,k\,\rangle_0 \,,
\end{eqnarray}
where $\langle\,k\,|\,\bar h_v\,i v\!\cdot\!D\,h_v\,|\,k\,\rangle_0=
v\cdot k$ and $\langle\,k\,|\,\bar h_v\,h_v\,|\,k\,\rangle_0=1$ are
the tree-level quark matrix elements. From (\ref{SeffPV}) we then
read off the renormalisation constants
\begin{eqnarray}\label{PVZs}
   Z_{\rm eff}^{-1} & = & 1 + {C_F\alpha_s\over\pi}\,\bigg(
    \ln{-2 v\cdot k\over\lambda} - 1 \bigg) \,,\nonumber\\
   Z_{v \cdot D\to \hat  1}^{-1} & = & -{C_F\alpha_s\over 2} \,.
\end{eqnarray}

In order to regulate higher power divergences, it is necessary to
introduce a more general form of the Pauli--Villars regularisation.
Instead of (\ref{eq:gpv1}), we shall use two subtractions and write
\begin{eqnarray}\label{eq:gpv31}
   G_{\mu\nu}(q) &=& -i g_{\mu\nu}\,\bigg( {1\over q^2}
    - {\gamma\over\gamma -1}\,{1\over q^2-\lambda^2}
    + {1\over\gamma-1}\,{1\over q^2- \gamma\lambda^2} \bigg)
    \nonumber\\
   &=& {-i g_{\mu\nu}\gamma\lambda^4
    \over q^2(q^2-\lambda^2)(q^2-\gamma\lambda^2)} \,.
\end{eqnarray}
Repeating the above calculation, we obtain
\begin{eqnarray}
   S_{\rm eff}^{-1}(k) &=& v\cdot k\,\Bigg\{ 1
    + {C_F\alpha_s\over\pi}\,\bigg[ \ln{-2 v\cdot k\over\lambda}
    + {\ln\gamma\over 2(\gamma-1)} \\
   &&\mbox{}+ {1\over\gamma-1}\,\Big( \gamma\sqrt{x^2-1}\,
    \arctan\sqrt{x^2-1} - \sqrt{x_\gamma^2-1}\,
    \arctan\sqrt{x_\gamma^2-1} \Big) \bigg] \Bigg\} \,, \nonumber
\end{eqnarray}
where $x_\gamma=\sqrt{\gamma}\,x=\sqrt{\gamma}\,\lambda/(-v\cdot k)$.
This leads to
\begin{eqnarray}
   Z_{\rm eff}^{-1} & = & 1 + {C_F\alpha_s\over\pi}\,\bigg(
    \ln{-2 v\cdot k\over\lambda} + {\ln\gamma\over 2(\gamma-1)}
    - 1 \bigg) \,,\nonumber\\
   Z_{v \cdot D\to \hat  1}^{-1} & = & -{C_F\alpha_s\over 2}\,
    {\sqrt{\gamma}\over\sqrt{\gamma}+1} \,.
\end{eqnarray}
Eqs.~(\ref{PVZs}) are recovered in the limit $\gamma\to\infty$.

The calculation of the matrix element of the kinetic energy operator
is more complicated. Consider again the diagrams in fig.~\ref{fig:1},
where now the shaded bubble represents the regulated propagator given
in (\ref{eq:gpv31}). We find that in the Feynman gauge only the
vertex and the tadpole diagrams give non-vanishing contributions. For
dimensional reasons the tadpole diagram is quadratically divergent.
Its contribution to  $S_{\rm kin}^{-1}(k)$ is
\begin{equation}
   -{3 C_F\alpha_s\over 4 \pi}\,\lambda^2\,
   {\gamma\ln\gamma\over\gamma-1} \,.
\end{equation}
However, this quadratic divergence is exactly cancelled by the
quadratic divergence of the vertex diagram. For the sum of all
one-loop contributions we find
\begin{eqnarray}\label{master}
   S_{\rm kin}^{-1}(k) &=& {C_F\alpha_s\over\pi}\,\Bigg\{
    {3\gamma\lambda^2\over\gamma-1}\,\Big(
    \sqrt{x_\gamma^2-1}^{-1}\arctan\sqrt{x_\gamma^2-1}
    - \sqrt{x^2-1}^{-1}\arctan\sqrt{x^2-1} \Big) \nonumber\\
   &&\mbox{}+ \Big[ k_\perp^2 - 3(v\cdot k)^2 \Big]\,\bigg[
    \ln{-2 v\cdot k\over\lambda} + {\ln\gamma\over 2(\gamma-1)} \\
   &&\mbox{}+ {1\over\gamma-1}\,\Big(
    \sqrt{x_\gamma^2-1}^{-1}\arctan\sqrt{x_\gamma^2-1}
    - \gamma\sqrt{x^2-1}^{-1}\arctan\sqrt{x^2-1} \Big) \bigg]
    \Bigg\} \,. \nonumber
\end{eqnarray}
Note that, in spite of the explicit factor of $\lambda^2$ in the
first term on the r.h.s., this term diverges only linearly with
$\lambda$.

We are now ready to compute the mixing of the kinetic energy operator
with lower-dimensional operators. We define
\begin{eqnarray}\label{eq:pv1}
   S_{\rm kin}^{-1}(k) &=& \langle\,k\,|\,\bar h_v\,(i D_\perp)^2
    h_v\,|\,k\,\rangle \nonumber\\
   \phantom{ \bigg[ }
   &=& Z_{\rm kin}^{-1}\,\langle\,k\,|\,\bar h_v\,(i D_\perp)^2
    h_v\,|\,k\,\rangle_0 + Z_{D_\perp^2\to (v\cdot D)^2}^{-1}\,
    \langle\,k\,|\,\bar h_v\,(i v\!\cdot\!D)^2 h_v\,|\,k\,\rangle_0
    \nonumber\\
   \phantom{ \bigg[ }
   &&\mbox{}+ Z_{D_\perp^2\to v\cdot D}^{-1}\,\lambda\,
    \langle\,k\,|\,\bar h_v\,i v\!\cdot\!D\,h_v\,|\,k\,\rangle_0
    + Z_{D_\perp^2\to\hat 1}^{-1}\,\lambda^2\,
    \langle\,k\,|\,\bar h_v\,h_v\,|\,k\,\rangle_0 \,.
\end{eqnarray}
{}From an expansion of (\ref{master}) in the limit of large
$\lambda$, we obtain
\begin{eqnarray}\label{Zresults}
   Z_{\rm kin}^{-1} & = & 1 + {C_F\alpha_s\over\pi}\,\bigg(
    \ln{-2 v\cdot k\over\lambda} + {\ln\gamma\over 2(\gamma-1)}
    \bigg) \,, \nonumber\\
   Z_{D_\perp^2\to (v\cdot D)^2}^{-1}
   &=& -{3 C_F\alpha_s\over\pi}\,\bigg(
    \ln{-2 v\cdot k\over\lambda} + {\ln\gamma\over 2(\gamma-1)}
    -1 \bigg) \,, \nonumber\\
   Z_{D_\perp^2\to v\cdot D}^{-1} &=& {3 C_F\alpha_s\over 2}\,
    {\sqrt{\gamma}\over\sqrt{\gamma}+1} \,, \nonumber\\
   Z_{D_\perp^2\to\hat 1}^{-1} &=& 0 \,.
\end{eqnarray}
As mentioned above, there is no quadratic divergence in the sum of
all diagrams, i.e.\ in Pauli--Villars regularisation there is no
mixing of the kinetic energy operator with the operator $\bar
h_v\,h_v$. We have checked that this result holds true in an
arbitrary number of space-time dimensions. Note that the coefficient
$Z_{\rm kin}^{-1}$ is related to the self-energy of the heavy quark
by
\begin{equation}\label{RPIrela}
   Z_{\rm kin}^{-1} = {{\rm d}\over{\rm d}(v\cdot k)}\,
   S_{\rm eff}^{-1}(k) \,.
\end{equation}
This relation is a consequence of reparametrisation invariance
\cite{luma}. It is satisfied also for the expressions given in
(\ref{Sfuns}), which were obtained in dimensional regularisation.

\subsection{Momentum flow regularisation}

We shall now discuss another regularisation scheme in which a hard UV
cut-off is used to compute the quark matrix elements in the HQET. It
is based on the observation that the resummation of renormalon chains
is equivalent to performing one-loop calculations with a running
coupling constant \cite{new}. Consider again the matrix element of
the operator $\bar h_v\,i v\!\cdot\!D\,h_v$. Imagine performing the
one-loop calculation of this matrix element with a running coupling
constant. In the Landau gauge, the result can be written in the form
\begin{equation}\label{taurep}
   S_{\rm eff}^{-1}(k) = v\cdot k\,\bigg\{ 1 +
   \int\limits_0^\infty\!{\rm d}\tau\,\widehat w_{v\cdot D}(\tau)\,
   {\alpha_s(\sqrt{\tau}\omega)\over\pi}
   \bigg\} \,,
\end{equation}
where we assume that $\omega=-2 v\cdot k\gg\Lambda_{\rm QCD}$. It has
been shown in ref.~\cite{new} that the distribution function
$\widehat w_{v\cdot D}(\tau)$ is related to the (unrenormalised)
Borel transform of the inverse propagator. One finds
\begin{equation}\label{wSrela}
   \int\limits_0^\infty\!{\rm d}\tau\,\widehat w_{v\cdot D}(\tau)\,
   \tau^{-u} = {3 C_F\over 2}\,{\Gamma(1-u)\,\Gamma(-1+2 u)\over
   \Gamma(2+u)} \,,
\end{equation}
which can be inverted to give
\begin{equation}
   \widehat w_{v\cdot D}(\tau) = {C_F\over 8\tau^2}\,
   \Big\{ (1+4\tau)^{3/2} - 1 - 6\tau \Big\} \,.
\end{equation}
The representation (\ref{taurep}) offers a natural way to introduce a
hard UV cut-off, since the product $\sqrt{\tau}\omega$ has the
interpretation of a physical scale. The distribution function
$\widehat w_{v\cdot D}(\tau)$ controls the momentum flow in the
one-loop self-energy diagram. For large values of $\tau$, this
function falls off proportional to $1/\sqrt{\tau}$, so that the
integral is linearly UV divergent. Introducing a hard UV
cut-off\footnote{In principle one should also introduce an IR
cut-off, since the running coupling constant is not well-defined in
the low-momentum region. This fact is irrelevant at one-loop order,
however, where one can neglect the running of the coupling constant.}
$\tau_{\rm UV}=\lambda^2/\omega^2$ and expanding the result for
$\lambda\to\infty$, we find for the renormalisation factors defined
in (\ref{eq:pv2})
\begin{eqnarray}
   Z_{\rm eff}^{-1} &=& 1 + {3 C_F\over 2}\,{\alpha_s\over\pi}\,
    \bigg( \ln{-2 v\cdot k\over\lambda} -  {1\over 2} \bigg)
    \,,\nonumber\\
   Z_{v \cdot D \to \hat  1}^{-1} &=& -{C_F\alpha_s\over\pi} \,.
\end{eqnarray}
Note the similarity to the results (\ref{PVZs}) obtained with a
Pauli--Villars regulator.

Let us now consider the quark matrix element of the kinetic energy
operator. We define
\begin{equation}
   S_{\rm kin}^{-1}(k) = k_\perp^2\,\bigg\{ 1
   + \int\limits_0^\infty\!{\rm d}\tau\,\widehat w_{D_\perp^2}(\tau)
   \,{\alpha_s(\sqrt{\tau}\omega)\over\pi} \bigg\}
   + (v\cdot k)^2\,\int\limits_0^\infty\!{\rm d}\tau\,
   \widehat w_{(v\cdot D)^2}(\tau)\,
   {\alpha_s(\sqrt{\tau}\omega)\over\pi} \,.
\end{equation}
Using again the relation between the distribution functions and the
Borel transforms given in (\ref{Sfuns}), we obtain
\begin{eqnarray}
   \widehat w_{D_\perp^2}(\tau) &=& -{3 C_F\over 16\tau^2}\,
    \Big( \sqrt{1+4\tau} - 1 \Big)^2 \,, \nonumber\\
   \phantom{ \bigg[ }
   \widehat w_{(v\cdot D)^2}(\tau) &=& -3\,
    \widehat w_{v\cdot D}(\tau) \,.
\end{eqnarray}
For large values of $\tau$, the first function falls off proportional
to $1/\tau$, so that the corresponding integral is logarithmically UV
divergent. After a straightforward calculation we find for the
renormalisation factors defined in (\ref{eq:pv1})
\begin{eqnarray}
   Z_{\rm kin}^{-1}&=& 1 + {3 C_F\over 2}\,{\alpha_s\over\pi}\,
    \bigg( \ln{-2 v\cdot k\over\lambda} + {1\over 2} \bigg) \,,
    \nonumber\\
   Z_{D_\perp^2 \to (v \cdot D)^2}^{-1} &=& -{9 C_F\over 2}\,
    {\alpha_s\over\pi}\,\bigg( \ln{-2 v\cdot k\over\lambda}
    - {1\over 2} \bigg) \,, \nonumber\\
   Z_{D_\perp^2 \to v \cdot D}^{-1} &=& {3 C_F\alpha_s\over\pi} \,,
    \nonumber\\
   Z_{D_\perp^2 \to \hat 1}^{-1} &=& 0 \,.
\end{eqnarray}
Once again we find that the quadratic divergence, which is expected
on dimensional grounds, is absent. In the momentum flow
regularisation scheme, the absence of the quadratic divergence is
directly linked to the absence of an UV renormalon pole at $u=1$ in
the Borel transform $\widetilde S_{\rm kin}^{-1}(k,u)$ in
(\ref{Sfuns}), since the Borel transform determines the distribution
function through an integral relation of the type (\ref{wSrela}).

It is instructive to note the close similarity of the above results
with (\ref{Zresults}). In both regularisation schemes and in an
arbitrary covariant gauge, the results for the renormalisation
factors can be written in the form
\begin{eqnarray}
   Z_{\rm eff}^{-1} &=& 1 + {C_F\alpha_s\over\pi}\,\bigg(
    {3-a\over 2}\,\ln{-2 v\cdot k\over\lambda} + c \bigg) \,,
    \nonumber\\
   Z_{v \cdot D\to \hat  1}^{-1} &=& -{C_F\alpha_s\over\pi}\,d \,,
\end{eqnarray}
and
\begin{eqnarray}
   Z_{\rm kin}^{-1} &=& \bigg( 1 + v\cdot k\,
    {{\rm d}\over{\rm d}v\cdot k} \bigg)\,Z_{\rm eff}^{-1} \,,
    \nonumber\\
   Z_{D_\perp^2 \to (v\cdot D)^2}^{-1} &=& -3 \Big(
    Z_{\rm eff}^{-1} - 1 \Big) \,, \nonumber\\
   Z_{D_\perp^2 \to v \cdot D}^{-1} &=& -3
    Z_{v\cdot D\to\hat 1}^{-1} \,, \nonumber\\
   Z_{D_\perp^2\to\hat 1}^{-1} &=& 0 \,,
\end{eqnarray}
where
\begin{equation}
   c_{\rm PV} = {\ln\gamma\over 2(\gamma-1)} - 1 \,,\qquad
   d_{\rm PV} = {\pi\over 2}\,{\sqrt{\gamma}\over
   \sqrt{\gamma} + 1}
\end{equation}
in Pauli--Villars regularisation (in the Feynman gauge), and
\begin{equation}
   c_{\rm MF} = -{1\over 2} \,,\qquad
   d_{\rm MF} = 1
\end{equation}
in the momentum flow regularisation (in the Landau gauge). The
coefficient $c$ is gauge dependent, but the constant $d$ is not.

\subsection{Lattice regularisation}
\label{subsec:lattice}

In this section we discuss the one-loop corrections to the propagator
and to the kinetic energy operator in the lattice regularisation. All
results reported in this section have been obtained in the static
theory ($\vec v=0$) in Euclidean space.

Among the possible lattice formulations of the HQET we choose the
simplest one, in which the heavy quark propagator in a background
gauge field $U$ has the form \cite{Eichten}
\begin{equation}\label{eq:s0}
   S_h^0(x,y) = \theta(x^4- y^4)\,\delta(\vec x-\vec y\/)\,
   {\cal P}_{\vec x}(x^4,y^4) \,,
\end{equation}
where ${\cal P}_{\vec x}(x^4,y^4)$ is the path-ordered exponential
from $(\vec x,x^4)$ to $(\vec x,y^4)$ along a path whose position in
space is constant:
\begin{equation}\label{eq:poe}
   {\cal P}_{\vec x}(x^4,y^4)\equiv U_4^\dagger(\vec x,x^4-a)\,
   U_4^\dagger(\vec x,x^4-2a)\dots U_4^\dagger(\vec x,y^4+a)\,
   U_4^\dagger(\vec x,y^4) \,.
\end{equation}
At one-loop order, the calculation of the heavy quark self-energy in
the Feynman gauge gives \cite{mms,eh1,blp}
\begin{equation}\label{eq:s0pert}
   \langle S^0_h(x,y)\rangle = \theta(x^4-y^4)\,
   \delta(\vec x-\vec y\/)\,\Big\{ 1 + \alpha_s(X t a^{-1} + Y)
   \Big\} + O(\alpha_s^2) \,,
\end{equation}
where $t=x^4-y^4$, $\alpha_s=g_0^2/4\pi$ is the bare lattice coupling
constant, and $\langle\dots\rangle$ denotes the average over the
field configurations. The quantities $X$ and $Y$ are given by
\begin{eqnarray}\label{eq:x}
   X &=& -{C_F\over 4}\int\limits_{-\pi}^\pi
    {{\rm d}^3k\over 2\pi^2}\,{1\over A} \,,\nonumber\\
   Y &=& {C_F\over 4}\int\limits_{-\pi}^\pi
    {{\rm d}^3k\over 2\pi^2}\,{1\over A\sqrt{(1+A)^2-1}} \,,
\end{eqnarray}
with
\begin{equation}
   A = \sum_{i=1}^{3} (1-\cos k_i) \,.
\end{equation}
The constant $Y$ is logarithmically divergent. It contributes to the
wave-function renormalisation and hence to the renormalisation of any
local operator containing the heavy quark field. The quantity $X$ is
gauge invariant and has to be identified with the coefficient of the
operator $\bar h_v\,h_v$. Adopting the notation introduced in
(\ref{eq:pv2}), where now $\lambda=1/a$ plays the role of the UV
cut-off, one obtains
\begin{equation}
   Z_{v \cdot D\to \hat  1}^{-1} = -\alpha_s\,X
\end{equation}
for the coefficient of the linearly divergent term.

The corrections to the kinetic energy operator can be computed from
\begin{equation}\label{eq:s1a}
   S_h^{1}(x,y) = \theta(x^4-y^4)\,\delta(\vec x-\vec y\/)\,
   \sum_{w^4=y^4}^{x^4} {\cal P}_{\vec x}(x^4,w^4)\,
   \vec D^2(\vec x,w^4)\,{\cal P}_{\vec y}(w^4,y^4) \,.
\end{equation}
For the operator $\vec D^2$ we take
\begin{equation}\label{eq:d2la}
   \vec D^2(\vec z,z^4)\,f(z) = {1\over a^2}\,\sum_{j=1}^{3}
   \Big( U_j(z)\,f(z+a\hat\jmath) + U^{\dagger}_j(z-a\hat\jmath)\,
   f(z-a\hat\jmath) - 2 f(z) \Big) \,,
\end{equation}
where $\hat\jmath$ denotes the unit vector in the $j$-direction. At
the tree level one finds
\begin{equation}\label{eq:od2}
   \langle S_h^{1}(x,y)\rangle_0 = \theta(x^4-y^4)\,
   \delta(\vec x-\vec y\/)\,{1\over a^2}\,\sum_{w^4=y^4}^{x^4}
   \sum_{j=1}^{3} \Big\{ \delta(\vec x+a\hat\jmath-\vec y\/)
   + \delta(\vec x-a\hat\jmath-\vec y\/)
   - 2\delta(\vec x -\vec y\/) \Big\} \,.
\end{equation}
At one-loop order one obtains \cite{mms}
\begin{eqnarray}\label{eq:1lsd2}
   \langle S_h^{1}(x,y)\rangle &=& \theta(x^4-y^4)\,
    \delta(\vec x-\vec y\/)\,\alpha_s\,(W t a^{-2} + 2 X a^{-1})
    \nonumber\\
   &&\mbox{}+ \Big[ \Delta + \alpha_s\,(X t a^{-1} + Y) \Big]\,
    \langle S_h^{1}(x,y)\rangle_0 \,,
\end{eqnarray}
where $X$ and $Y$ are defined in (\ref{eq:x}), $\Delta = 1+
\frac{1}{6}\,\alpha_s W$, and
\begin{equation}\label{eq:w}
   W = -C_F\,\int\limits_{-\pi}^\pi {{\rm d}^3k\over 2\pi^2}\,
   {1\over\sqrt{(1+A)^2-1}} \,.
\end{equation}
Using the notation introduced in (\ref{eq:pv1}), we find
\begin{eqnarray}
   Z_{D_\perp^2\to v \cdot D}^{-1} = 2\alpha_s\,X \,,\nonumber\\
   Z_{D_\perp^2\to \hat  1}^{-1} = - \alpha_s\,W \,.
\end{eqnarray}
Thus, in this case the mixing coefficient of $\bar h_v\,(i D_\perp)^2
h_v$ with $\bar h_v\,h_v$ does not vanish. Moreover, since
reparametrisation invariance is broken on the lattice, the
renormalisation of the kinetic energy operator differs from the wave
function renormalisation by a factor $\Delta\ne 1$.

\section{Conclusions}
\label{sec:6}

In the HQET the coefficients of higher-order terms in the expansion
in inverse powers of the heavy quark mass are proportional to matrix
elements of higher-dimensional operators. In general these matrix
elements diverge in perturbation theory as powers of the UV cut-off,
and using dimensional regularisation their perturbation series are
not Borel summable due to the presence of UV renormalon singularities
in the Borel plane. The ambiguities due to the presence of these
renormalons are cancelled by those due to IR renormalons in the
Wilson coefficient functions of lower-dimensional operators.

In this paper we have investigated in some detail the relation
between power divergences and renormalons. This correspondence, which
is based on dimensional arguments, appears to fail in one important
case, that of the kinetic energy operator. In a lattice
regularisation $\bar h_v\,(i D_\perp)^2 h_v$ mixes with $\bar
h_v\,h_v$ with a quadratically divergent coefficient, as expected
from the naive degree of divergence of the corresponding Feynman
diagrams. However, at least to one-loop order in perturbation theory,
this quadratic divergence is absent when using the Pauli-Villars and
momentum flow regularisations. Moreover, the corresponding UV
renormalon is absent using dimensional regularisation. The absence of
the corresponding IR renormalon in the pole mass had already been
noted in ref.~\cite{bb}. These results suggest that there may be some
symmetry, which prevents $\bar h_v\,(i D_\perp)^2 h_v$ from mixing
with $\bar h_v\,h_v$, and that this symmetry is broken with the
lattice regularisation. An obvious candidate for such a symmetry is
Lorentz (or Euclidean) invariance, which takes the form of
reparametrisation invariance in the effective theory
\cite{luma,cino}. This symmetry does not seem to forbid such a
mixing, however. Given the relevance of the kinetic energy operator
to studies of the spectroscopy and inclusive decays of heavy hadrons
\cite{spectrum}--\cite{inclu4}, it is important to understand whether
the absence of the power divergences and the corresponding renormalon
is an accident of one-loop perturbation theory and the bubble
approximation, or whether it is a consequence of a more general
principle. We hope that this paper will stimulate further
investigation of this puzzle.

\section*{Acknowledgements}

We thank M.~Beneke and J.~Nieves for interesting discussions. G.M.\
acknowledges partial support from M.U.R.S.T., and C.T.S.\
acknowledges the Particle Physics and Astronomy Research Council for
its support through the award of a Senior Fellowship. G.M.\ and
C.T.S.\ acknowledge partial support by the EC contract
CHRX-CT92-0051.

\newpage

\centerline{\Large\bf Figure}

\end{document}